\documentclass[conference]{IEEEtran}
\IEEEoverridecommandlockouts
\usepackage{cite}
\usepackage{amsmath,amssymb,amsfonts}
\usepackage{algorithmic}
\usepackage{graphicx}
\usepackage{textcomp}
\usepackage{hyperref}
\usepackage{float}
\usepackage{xcolor}
\usepackage[T1]{fontenc}
\usepackage[utf8]{inputenc}
\usepackage{todonotes}
\def\BibTeX{{\rm B\kern-.05em{\sc i\kern-.025em b}\kern-.08em
    T\kern-.1667em\lower.7ex\hbox{E}\kern-.125emX}}

\usepackage[many]{tcolorbox}
\newtcolorbox{colorb}{
enhanced,
boxrule=0pt,frame hidden,
borderline west={2pt}{0pt}{green!50!black},
colback=green!05!white,
sharp corners
}

\begin{document}

\title{On Developing an Artifact-based Approach to Regulatory Requirements Engineering}

 \author{
    \IEEEauthorblockN{
        Oleksandr Kosenkov*, Michael Unterkalmsteiner \\ 
        Daniel Mendez*, 
        Davide Fucci,
        Tony Gorschek}
    \IEEEauthorblockA{\textit{Blekinge Institute of Technology}\\
        Karlskrona, Sweden \\
        \{firstname\}.\{lastname\}@bth.se}
    \and
    \IEEEauthorblockN{
        Jannik Fischbach}
    \IEEEauthorblockA{\textit{Netlight Consulting GmbH and *fortiss GmbH}\\
        Munich, Germany \\
        jannik.fischbach@netlight.com}
}

\maketitle

\begin{abstract}
\emph{Context:} Regulatory acts are a challenging source when eliciting, interpreting, and analyzing requirements. Requirements engineers often need to involve legal experts who, however, may often not be available. This raises the need for approaches to regulatory Requirements Engineering (RE) covering and integrating both legal and engineering perspectives.
\emph{Problem:} Regulatory RE approaches need to capture and reflect both the elementary concepts and relationships from a legal perspective and their seamless transition to concepts used to specify software requirements. No existing approach considers explicating and managing legal domain knowledge and engineering-legal coordination.
\emph{Method:} We conducted focus group sessions with legal researchers to identify the core challenges to establishing a regulatory RE approach. Based on our findings, we developed a candidate solution and conducted a first conceptual validation to assess its feasibility.
\emph{Results:} We introduce the first version of our Artifact Model for Regulatory Requirements Engineering (AM4RRE) and its conceptual foundation. It provides a blueprint for applying legal (modelling) concepts and well-established RE concepts. Our initial results suggest that artifact-centric RE can be applied to managing legal domain knowledge and engineering-legal coordination.
\emph{Conclusions:} The focus groups that served as a basis for building our model and the results from the expert validation both strengthen our confidence that we already provide a valuable basis for systematically integrating legal concepts into RE. This overcomes contemporary challenges to regulatory RE and serves as a basis for exposure to critical discussions in the community before continuing with the development of tool-supported extensions and large-scale empirical evaluations in practice.
\end{abstract}


\section{Introduction}
\textbf{Context} Regulatory Requirements Engineering (RE) aims to interpret regulatory acts (any written, public, official, obligatory source of norms) and infer software requirements. This is crucial as the number and complexity of regulatory acts applicable to software systems have been growing in recent years~\cite{wagner2004software, arnould2021complexity}. Regulatory RE activities aim at producing a consistent interpretation of the law~\cite{otto2007addressing, boella2014critical} and meet the expectations of legal stakeholders (regulators and legal experts)~\cite{alexander2004understanding, oberle2012engineering}. In practice, there is a disparity between available RE approaches and the expectations of legal stakeholders. For example, all applicable regulatory acts must be considered throughout the entire software development life cycle; yet, existing approaches usually focus on one specific regulatory act or process area only in isolation~\cite{zdun2012guest}. To be practically applicable, regulatory RE approaches should satisfy the demands of legal practice and cover all the activities involved~\cite{boella2014critical}. 
Whatever approach is eventually chosen to engage in regulatory RE, it needs to incorporate legal knowledge in some form~\cite{boella2013managing}.

\textbf{Problem} Extensive work was conducted on processing and analyzing regulatory acts in RE and AI\&Law research. Nevertheless, some challenges remain unresolved as there is still a ``gap between engineers and legal experts''~\cite{mubarkoot2022software}. In research that considers the automatic processing of regulatory acts without legal experts' involvement, there is no agreement about the semantic concepts that can be automatically extracted from regulatory acts~\cite{sleimi2021automated}. Available studies corroborate that engineering-legal interactions are challenging~\cite{bobkowska2010efficient} because of the differences in language and terminology~\cite{bobkowska2010efficient}, the simplistic perception of each domain in regulatory RE~\cite{boella2014critical, hjerppe2019general}, and the implicit nature of legal knowledge during the interaction~\cite{torre2021modeling}. Taking into account that it is one of the important efforts in RE to address and reconcile different stakeholders' viewpoints in RE~\cite{pouloudi1999aspects}, we formulate the following main objective to address contemporary challenges:

\emph{We aim at developing the first version of a regulatory RE approach that satisfies the demands of legal experts and supports engineering-legal coordination.}

\textbf{Contribution} In this paper, we report on three basic contributions: (1) focus group research with legal researchers as a ground for prioritizing core challenges related to capturing and reflecting legal concepts in regulatory RE artifacts. Based on the results, we (2) developed the first version of the Artifact Model for Regulatory Requirements Engineering (AM4RRE) that describes the typical work products, their contents, and their dependencies as required in regulatory RE. Finally, we report on (3) a conceptual validation with requirements engineering and legal researchers, providing a first insight into the feasibility of our approach from legal and requirements engineering perspectives. Based on the results of our conceptual validation, we hypothesize that artifact-based regulatory RE has the potential to (1) model legal domain knowledge to a certain degree by accurately capturing and reflecting the elementary legal concepts and their relationships and (2) support interdisciplinary engineering-legal coordination. We plan to validate these hypotheses in future studies.

\textbf{Relation to existing work} To the best of our knowledge, no previous studies reported the benefits of the application of artifact-based and model-based RE for managing domain knowledge and coordination of roles involved in regulatory RE. We report our study results and outline future research plans to get feedback and trigger community discussion.

\section{Terminology}
Regulatory acts are written sources for functional and non-functional software requirements. A \emph{regulatory act} is any written, public, official, obligatory source of norms (incl. laws, directives, etc.)~\cite{levi2011regulation,brownsword2009law}. The \emph{legal terms} used in regulatory texts have both a linguistic and legal meaning~\cite{husa2012understanding,waltl2019semantic} (e.g., term ``data subject'' in GDPR). It is important to distinguish between the layman's interpretation of a term and its meaning narrowed down by legal definitions. \emph{Legal concepts} are ``stereotypical'' semantic elements of regulatory acts that aggregate minimally meaningful legal meaning for legal interpretation and reasoning and have an impact on their outcomes~\cite{ross1957tu, ashley2003predictive, sartor2009legal, al2015factors} (for example, ``data subject'' is a term which is a specification of the legal concept ``legal subject'').
\emph{Legal interpretation} is the process applied by legal experts that uses legal domain knowledge to assign meaning to the regulatory text in a particular case in which there is a doubt about an appropriate meaning of a regulatory act~\cite{wroblewski1969legal,soames2011toward,cheng2012legal}. In our work, we consider legal interpretation as an activity conducted in RE to infer software requirements and not as an ``operative interpretation'' conducted only by authorities~\cite{wroblewski1969legal,soames2011toward}. For this study, we define \emph{legal knowledge} as knowledge about legal concepts, their relationships and the ways for its application for legal interpretation. Based on existing literature, we conclude that \emph{regulatory RE} is the process of deriving software requirements from intentionally abstract regulatory texts by applying legal interpretation and legal concepts in the context of a specific software project.

Regulatory RE, in many cases, can require \emph{engineering-legal coordination}, which can be understood as the exchange of information or another form of interaction between software engineering and legal expert roles in processing legal and RE artifacts. Coordination implies that roles work on the same or related artifacts and/or that the roles are mutually dependent on each other's contributions.

\section{Related Work}\label{sec:relatedWork}
\textbf{Conceptual models} Extensive work for identifying legal concepts and semantic metadata in regulatory texts has been done in both RE and AI\&Law research. Nevertheless, previous research was mainly focused on the identification of general linguistic concepts (e.g., ``Actor'', ``Actions'')~\cite{massey2010evaluating, boella2013managing, ingolfo2014nomos, zeni2015gaiust, engiel2017tool, sleimi2021automated}, rather than legal concepts specific to regulatory texts. This is despite the role of legal concepts being well-known. For example, most work in the area of automatic processing of regulatory texts is based on the work of Hohfeld on the fundamental legal conceptions dating back to 1917~\cite{sleimi2021automated}. Nevertheless, the work by Hohfeld remains the only legal work \emph{widely and thoroughly} considered. Under-consideration of the legal perspective could be one of the reasons for the missing consensus about the types of legal concepts that are useful for analyzing regulatory acts~\cite{sleimi2021automated}.

\textbf{Legal interpretation for elicitation of requirements from regulatory acts} Many publications have suggested a structured approach for the elicitation of requirements from regulatory acts~\cite{massey2010evaluating, bobkowska2010efficient, oberle2012engineering, beckers2012integrated, boella2013managing, raykar2021iterative}. Some studies in RE consider legal interpretation as a required part of the elicitation of requirements from regulatory acts and at least briefly describe it~\cite{breaux2007systematic, oberle2012engineering, boella2013managing, muthuri2016argumentation}. Nevertheless, even works specifically focusing on exploring legal interpretation~\cite{muthuri2016argumentation} have suggested more general stages of the process (domain classification, argumentation structure, etc.). In some contributions, the authors suggested approaches that resemble legal interpretation without identifying it as such (e.g.,\cite{raykar2021iterative}). Overall, however, existing literature treats the legal interpretation at a rather abstract level that we deem insufficient for application in practice. Related research assumes so far that there is no one logical framework for legal interpretations, in addition to numerous contradictions and inconsistencies on how legal interpretation is eventually conducted~\cite{devenish1991nature}. Other publications also considered interpretation executed by legal experts as ``eclectic''~\cite{oberle2012engineering}.

\textbf{Engineering-legal coordination}\label{sec:interaction} Multiple publications have acknowledged the need for the interaction of software engineers with legal experts for the elicitation of requirements from regulatory acts (e.g., for the identification of the applicable regulatory acts, mapping of terms)~\cite{massey2010evaluating,bobkowska2010efficient,oberle2012engineering,beckers2012integrated,boella2013managing,engiel2017tool,raykar2021iterative}. Nevertheless, the description of the interaction was usually provided at an abstract level as well, with little to no guidance on how to do it. Existing literature in software engineering and AI\&Law points to the need to capture and reflect the elementary concepts and relationships from a legal perspective. Nevertheless, to the best of our knowledge, there is no systematic approach to that available yet and our work presented here aims at closing this gap.

\section{Methodology overview}\label{sec:methodology}
To achieve the aim of our research and develop an RE approach that captures the legal knowledge and supports engineering-legal coordination, we have formulated the following research questions:

\begin{itemize}\setlength{\itemindent}{.2in}
    \item[\textbf{RQ 1}] What are the challenges in regulatory RE from the perspective of legal researchers?
    \item[\textbf{RQ 2}] How can the challenges in regulatory RE be effectively addressed in an integrative regulatory RE approach?
    \item[\textbf{RQ 3}] How do RE and legal researchers assess the applicability of such an integrative regulatory RE approach?
\end{itemize}

In particular, we aim to create a conceptual basis for our regulatory RE approach and develop the first version of such an approach to validate such conceptual foundation. In the remainder of this section, we discuss the research methods used to answer those questions.

\subsection{Focus group design and implementation}
To address RQ 1, we conducted focus group research following the guidelines by Kontio et al.~\cite{kontio2004using}. 
Specifically, we involved four legal researchers in two focus groups. Our aim was to explore the core challenges in regulatory RE and generate ideas to address these challenges in a way that is valid from the legal perspective.
Considering that RQ 1 is exploratory, for the first focus group with two legal researchers, we conducted multiple sessions (1 to 1.5 hours long) until saturation was reached, with no new major challenges and ideas.
The first author of the paper acted as a facilitator of the discussions: (1) introducing the legal researchers to approaches in RE (depending on the discussion context), (2) assessing if the results of the focus group discussions are on the required level of abstraction, (3) asking additional clarifying questions, and (4) summarizing the discussions after each session. In some sessions, a virtual whiteboard was used to structure the ideas and discuss complex questions. The first author applied thematic analysis to the notes he made or provided by participants after each focus group session to analyse the results. If any new important themes were discovered in the later sessions, we conducted additional analysis of materials generated in previous sessions. From our experience, some of the insights (for example, about legal concepts and purposes of legal interpretation) were scattered across all the focus group sessions and must be collected and integrated. 
The sessions' structure, scope, and objectives were planned based on the results from the previous sessions. Eventually, five focus group sessions were conducted, which covered four following topics: (1) the purpose of regulatory RE, existing challenges reported in the literature, and state of the art in regulatory RE methods (2 sessions); (2)  legal domain and potential approaches to model it; (3) legal interpretation process; (4) artifacts available when conducting a legal interpretation. As a starting point for the first session, we used intermediary results of an ongoing systematic mapping study. We discussed four main categories of challenges in regulatory RE reported in the literature: (1) abstract and vague nature of regulatory acts (e.g.,~\cite{ayala2018grace}), (2) changeability of regulatory acts (e.g.,~\cite{hjerppe2019general}), (3) demand for legal domain knowledge (e.g.,~\cite{peixoto2020understanding}), and (4) evolution/changes of software context (e.g.,~\cite{dias2020perceptions}). After each session, the first author and legal researchers analyzed the results and read additional literature to prepare for the next session. Additional publications in regulatory RE were provided to the legal researchers on demand by the first author. In some cases, legal researchers identified and read such publications independently. There was no particular schedule for the focus group sessions. Rather, they were held as soon as the focus group participants were available, and the results of the previous session were analyzed. Overall, the sessions took place over twelve months. As a result of the focus group sessions with the first group of legal researchers, we formulated four challenges. After that, we conducted another one-hour-long focus group with two other legal researchers and discussed the main findings of the first focus group. The results of the second focus group were consistent with the initial findings, and we were able to extend the findings with additional details. The challenges we identified are presented in section~\ref{sec:focusGroups}

\subsection{Synthesis and model generation}
To answer RQ 2, the first author synthesized the results of the focus group sessions and reviewed further legal literature to develop a candidate solution described in Section~\ref{sec:artifactModel}. Legal researchers did not take part in the development of the model. Our synthesis indicates that legal concepts are insufficiently considered in regulatory RE methods. Simultaneously, the Hohfeldian system of jural relations is essential for existing regulatory RE methods~\cite{sleimi2021automated}. Driven by this finding, we searched the legal literature for other legal theories or approaches that can be adapted for regulatory RE purposes. We found such an approach to the core legal concepts in the work of Radbruch~\cite{radbruch2003rechtsphilosophie}. After comparing this work to the works of Hohfeld and Alexy, we selected Radbruch's approach as a basis for our approach. Further, we amended the concepts described by Radbruch with other concepts found in the literature related to the area of requirements engineering for GDPR compliance (e.g., information about the ``delegation'' concepts was found in EDPB Guidelines 07/2020). Definitions of the core concepts are available in open material package~\ref{sec:data}.

\subsection{Conceptual Validation}
We conducted a conceptual validation to validate the applicability of our approach (AM4RRE) and answer RQ3. Conceptual validation focuses on the validation and qualitative evaluation of the model's underlying theory~\cite{rao1998development}. In our case, the overall theory requiring validation was the applicability of artifact orientation to capture legal knowledge (in the form of legal concepts and their relationships) and support engineering-legal coordination independently of a particular regulation or field of the regulation (e.g., personal data protection). We did not use any particular regulation for a walk-through to achieve the validation purposes. Also, we involved researchers who could consider applying such a model across different regulations and projects.

Our validation involved five participants: two RE researchers (RER1, RER2) who have experience with regulations in the context of RE, one legal informatics researcher (LR1) with both computer science and legal background and two legal researchers (LR2, LR3) experienced in collaborations in software engineers contexts. The legal researchers from the focus group were not involved in the validation. To reduce the potential for bias, we selected the validation participants among the researchers with whom the first author did not have any research collaboration.
Our approach extends an original artefact-based model for RE (AMDiRE, more details in Sect.~\ref{sec:artifactModel}) with legal concepts. We validated our integrated approach via a walk-through of the original model and an explanation of the new model components. The study's first author conducted a validation of the model itself rather than of potential instances to focus on the principle applicability while remaining independent from and not getting lost in details of selected legal concepts important to chosen regulatory acts only. The participants were asked questions throughout the walk-through (see the detailed list of questions in the open science package~\ref{sec:data}). The questions were asked gradually during the walk-through and at the end of the walk-through. The questions asked during the walk-through mainly focused on the sufficiency of the elements of the models for conducting regulatory RE. The questions asked at the end concerned the model's capability to (1) explicate and compensate for legal knowledge and (2) support engineering-legal coordination. In addition, we discussed the potential benefits and drawbacks of the model in case of its practical application. We conducted five validation sessions involving one researcher in each (average session time was 1 hour).

\subsection{Threats to Validity}
One important threat to validity emerges from the interdisciplinary context where legal experts might, for example, have introduced bias. We have mitigated this threat in various ways. We based the discussions of focus groups on existing literature. We also considered concrete cases of regulatory RE and compliance to derive and discuss universally applicable ideas and challenges in focus groups. Further, the moderating first author has a background, experience, and degrees in law and engineering. Another important threat is that only two groups of four legal researchers were involved in focus groups. However, while involving further groups might have been possible, the saturation of the discussion outcomes (when no new major challenges and ideas emerged) strengthened our confidence that the results were suitable to begin our work under RQ2. Also, the results of the focus groups were in line with some of the findings in existing studies. We added a conceptual validation involving five participants to compensate for potential threats emerging from the sample size. Finally, this gives us the opportunity to further revise our model before beginning long-term implementations of tool support that we deem necessary for large-scale industrial case studies (in this sensitive topic area).

\section{Focus Group Results (RQ1)}\label{sec:focusGroups}
We answer RQ1 by summarizing the results of the focus group research. We report four challenges (Ch1 - Ch4) to capture and reflect the elementary legal concepts in regulatory RE methods.

\subsection{Ch1: Detachment from legal interpretation practice}
According to legal researchers, methods for eliciting requirements from regulatory acts that focus on the analysis of regulatory texts in isolation from the real-world context (such as in~\cite{ghanavati2014goal, sleimi2021automated, ingolfo2014nomos}) are only partially relevant from a legal perspective. \emph{Legal researchers suggested that regulatory RE approaches should incorporate or consider the legal interpretation process usually applied by legal experts in practice.} Otherwise, the logic behind the approach applied to process regulatory acts can be unclear to legal experts. Application of legal interpretation as in legal practice should help address heterogeneity and complexity of regulatory context (e.g., organization of regulatory acts according to fields of law, each of which has its own peculiarities).
One of the legal researchers denoted legal interpretation as a ``social practice'' rather than a strict process (this idea is also supported by literature~\cite{cheng2012legal}). The main purpose of legal interpretation is to ensure the achievement of high-level legal goals of regulatory acts and law. Under the changes in the social environment, the ways to achievement of such goals can change. Our further literature review confirmed that principles~\cite{ashley2009ontological}, social purposes~\cite{berman1993representing}, values and theories~\cite{bench2003model} are considered important in legal interpretation and reasoning.

\subsection{Ch2: Non-linear and iterative nature of legal interpretation}
Some regulatory RE approaches (such as~\cite{massey2010evaluating, bobkowska2010efficient}) see legal interpretations as linear. After applicable regulatory acts are identified, concrete regulatory acts are interpreted as iterative consideration of both regulatory acts and information about a case/system. \textit{According to legal researchers, in practice, the flow of the processing of both regulatory acts and information about the case they are applied to depends on the information processed.} For example, the discovery of new relevant properties of software systems in requirements by legal experts (e.g., particular types of processed data) could very well lead to identifying other applicable regulatory acts. It is thus a continuous back and forth along the engineering process where RE and design decisions and consideration of alternative compliance options may require new legal interpretation.
Previous studies also emphasized variability and complex interdependencies between different tasks as challenges for granular modeling of legal interpretation as an activity~\cite{devenish1991nature,oberle2012engineering}).

\subsection{Ch3: Ignorance of the software context}
Legal researchers have emphasized that law per se is usually not focused on software systems only. Regulatory acts are primarily concerned with human subjects and social relations rather than software systems~\cite{breaux2008analyzing}. Regulatory acts applicable to organizations or stakeholders are also indirectly relevant to software systems. This makes it essential to include a broader organizational and social context in regulatory RE.
Related work also highlights that compliance to regulatory acts exists both on system, process, and organizational levels~\cite{turetken2012capturing} and that regulatory RE should account for the context of the system under development~\cite{martin2018methods, zimmermann2020automation}.

\subsection{Ch4: Limited application of legal concepts in regulatory RE approaches}
Together with legal researchers, we have jointly analyzed and discussed concepts suggested in work by Sleimi~\cite{sleimi2021automated} that summarizes 11 different conceptual models. General concepts (e.g., ``agent'') that were identified in such conceptual models are understandable to legal experts, but they do not help to structure and analyze regulatory text in a way that legal experts and researchers would do. Legal experts analyze legal texts using their legal domain-specific concepts and by splitting text into elements using their legal knowledge and experience. Applying a different set of concepts for regulatory RE is a daunting task that may not produce the same results. Overall, our focus group sessions allowed us to identify challenges to modeling the elementary concepts and relationships from a legal perspective in regulatory RE approaches.

\begin{colorb}
\textbf{Answer to RQ1:} Our identified challenges mainly include the questions of how to apply legal domain knowledge (legal interpretation process, legal concepts) and how to consider the context of software systems explicitly.
\end{colorb}

\begin{figure}
\centerline{\includegraphics[width=1\linewidth]{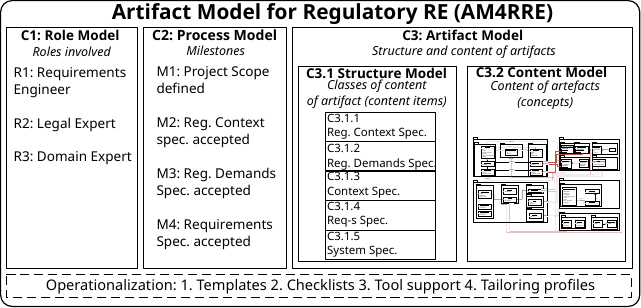}}
\caption{Overview of main components of the AM4RRE model}
\label{fig:am4rreComponents}
\end{figure}

\section{Artifact model for Regulatory Requirements Engineering - AM4RRE (RQ2)}\label{sec:artifactModel}
In this section, we describe our candidate solution for regulatory RE. We specifically opted for an implementation following an artifact-oriented philosophy over a process-oriented one~\cite{mendez2010meta} to address the challenges discussed under RQ1. 
In a nutshell, when following an artifact-oriented approach in RE, we define an artifact model of all results of RE (\textit{What}) as a backbone, rather than concrete RE activities and complex processes (\textit{How}), thus coping with the inherent variability in the interpretation process and, instead, focusing on core concepts and their dependencies. We define a content model that abstracts from modeling concepts and their relationships and a structure model that provides structure to the content model (comparable to a document outline as an overlay). The latter serves as an interface to the surrounding role model and process model (see also~\cite{mendez2010meta} for further details on artifact orientation). Choosing this approach helps us capture and model (and integrate) both legal and RE concepts, which can stay implicit for legal experts and requirements engineers in the case of process-oriented RE approaches. We use and extend the so-called Artifact model for Domain-independent Requirement Engineering (AMDiRE)~\cite{mendez2015artefact} as a state-of-the-art model to build on and extend it into our regulatory RE variant, called AM4RRE.
The basic components of the AM4RRE approach are as follows:
\begin{itemize}
	\item the role model (C1 in Figure~\ref{fig:am4rreComponents}) defines the roles (R) involved in regulatory RE and the responsibilities of each role;
	\item the process model (C2 in Figure~\ref{fig:am4rreComponents}) describes the main milestones (M) that can be used to define a specific process and workflow for regulatory RE;
	\item the artifact model (C3) captures the structure (C3.1 in Figure~\ref{fig:am4rreComponents}) of specification of both regulatory acts (C3.1.1, C3.1.2) and requirements (C3.1.3, C3.1.4, C3.1.5) and content (C3.2 in Fig.~\ref{fig:am4rreContent1}) of both regulatory acts and requirements.
\end{itemize}
Due to the space limitations in this paper, we briefly describe C1 and C2 and focus on the artifact model (C3).

\subsection{Role Model (C1)}
The role model (C1) of AM4RRE contains three roles that are involved in the regulatory RE process.
\begin{itemize}
    \item\emph{Requirements engineer (R1)}.
    \item\emph{Legal expert (R2)}.
    \item\emph{Domain expert (R3)}.
\end{itemize}

\textbf{Challenges addressed (RQ1)} The extension of the original AMDiRE model with the R2 legal expert and R3 domain expert allows for the \emph{application of legal norms to a particular case (Ch1)} and \emph{enables a more flexible exchange of information between requirements engineers and legal experts (Ch2)}. The inclusion of the role of a domain expert addresses the need for the \emph{consideration of software context (Ch3)}.

\subsection{Process Model (C2)}
Following the notion of artifact orientation, the process model (C2) guides the processing of regulatory acts by describing milestones in relation to the artifacts being processed. We include the following milestones:
\begin{itemize}
    \item M1: Project Scope defined.
    \item M2: Regulatory Context Specification accepted.
    \item M3: Regulatory Demands Specification accepted.
    \item M4: Requirements Specification accepted.
\end{itemize}

We identified these milestones based on the three core activities in the legal interpretation:  (1) identification of applicable regulatory acts (based on the project scope), (2) analysis of texts of applicable regulatory acts, and (3) derivation of context-specific requirements from regulatory acts.

\textbf{Challenges addressed (RQ1)} The process model was developed in a way to \emph{embrace the process of legal interpretation as suggested by legal researchers (i.e., iterative consideration of regulatory acts and software system) (Ch1)} and \emph{enable flexibility in the processing of the information required for legal interpretation (Ch2)}.

\subsection{Artifact Model (C3)}
The artifact model (C3) is the core component of the AM4RRE approach and describes the structure and contents of the artifacts specified in regulatory RE. 
The artifacts in the original artifact model AMDiRE~\cite{mendez2015artefact} are ordered along three levels of abstraction (``layers''), each capturing a particular type of artifact: context specification on a context layer (C3.1.3), (user) requirements specification on the requirements layer (C3.1.4), and system specification on the system layer (C3.1.5). To specify regulatory acts as a part of the context specification, we introduce two additional layers of abstraction. At these levels of abstraction, original legal concepts are specified to make legal knowledge and further interpretation of legal concepts explicit: 
\textit{Regulatory context layer} (C3.1.1) at which legal concepts contain meta-information determining interrelations between regulatory acts. Legal concepts belonging to this level of abstraction describe the properties of each regulatory act in relation to other regulatory acts. Hence, these properties are also inherited by concepts on the regulatory demands level. For example, Articles 2 and 3 of GDPR describe the jurisdiction of GDPR. This applicable jurisdiction applies to the concepts on the regulatory acts' structure level, i.e., concepts in Article 15 are applicable within the determined jurisdiction of GDPR. Regulatory context specification (as an intermediary result artifact of regulatory RE) captures these concepts that determine the applicability of different regulatory acts. Acceptance of the regulatory context specification marks milestone 2 (M2).
\textit{Regulatory demands layer} (C3.1.2) contains legal concepts describing the content of regulatory acts (rules that need to be adhered to). Regulatory demands specification capture concepts on this level of abstraction, which is also an intermediary result of regulatory RE. Acceptance of the regulatory demands specification marks the Milestone 3 (M3).

\begin{figure*}
\centerline{\includegraphics[width=0.6\linewidth]{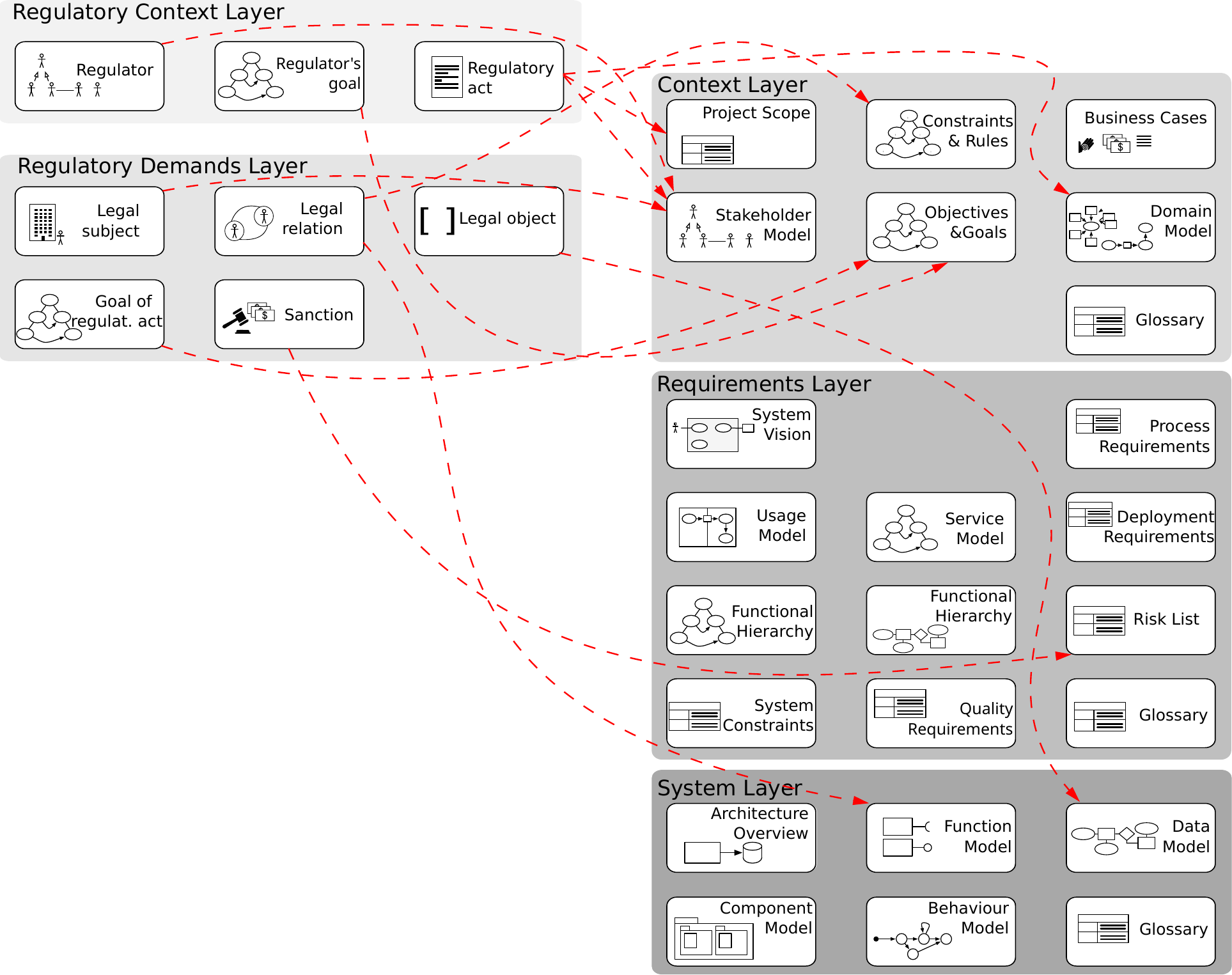}}
\caption{Pictogrammatic overview of the AM4RRE Artifact Model showing only the integration of legal concepts with requirements concepts (red associations).}
\label{fig:am4rreContent1}
\end{figure*}

Following the principles of artifact orientation, requirements concepts and legal concepts are represented in the artifact model with two views (see Fig.~\ref{fig:am4rreComponents}). 
The structure view (C3.1) provides structure by describing a taxonomy of content items, i.e. classes of the concepts that constitute the content of artifacts (for example, content item ``Regulator''). These elements also serve as interfaces to the process model (milestones) and the role model. The underlying content model (C3.2 visualized in a simplified manner in Figure~\ref{fig:am4rreContent1}) defines for each of the content items a blueprint to specify its contents by abstracting from the modeling concepts used to syntactically describe (1) concepts (for example, the legal concept of ``Jurisdiction'' belonging to content item ``Regulator'') as well as (2) relationships between the concepts, thus, facilitating syntactic consistency.

\begin{figure}
\centerline{\includegraphics[width=1\linewidth]{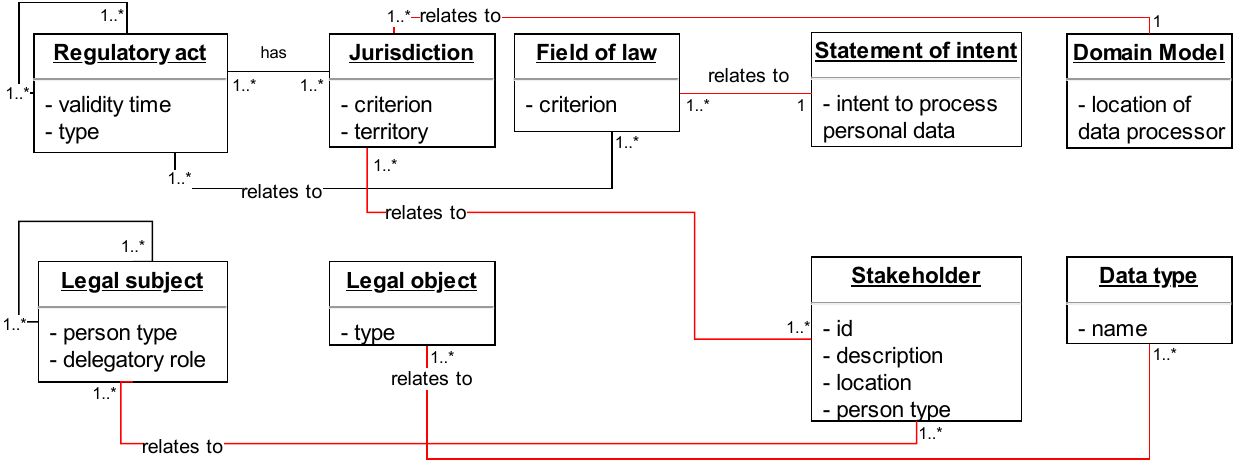}}
\caption{Fragment of the AM4RRE content model (C3.2) depicting concepts (classes); associations remain simplified.}
\label{fig:am4rreContent2}
\end{figure}

Relationships between legal concepts and requirements concepts (shown as red lines between classes in Figure~\ref{fig:am4rreContent2}) guide the interpretation of legal concepts with requirements concepts. These relationships are instrumental for mapping the concepts and ``translating'' legal concepts to achieve Milestone 4 ``Requirements Specification accepted'' (once requirements are derived from regulatory acts). These relationships were established based on our analysis, i.e. they require further refinement and evaluation via case study research that is in the scope of our future work.

\textbf{Challenges addressed (RQ1)} The artifact model was developed based on concepts applied in the legal domain by legal researchers and experts to overcome challenges of \emph{limited application of legal concepts (Ch4)}, which allowed to make AM4RRE \emph{closer to the legal practice of interpreting regulatory acts (Ch1)}. Also, AM4RRE contains context layer requirements artifacts (such as stakeholder model, domain model) that are crucial to \emph{address regulatory acts imposed on the context of software systems (Ch3)}. Another benefit of AM4RRE is the explication of legal knowledge in regulatory context specification and a regulatory demands specification.

\begin{colorb}
\textbf{Answer to RQ2:} Artifact-based RE can be adapted to capture legal knowledge and support engineering-legal coordination. This can be achieved by explicating legal knowledge in RE artifacts and structuring the interaction between the roles involved in regulatory RE.
\end{colorb}

\section{Conceptual Validation Results (RQ3)}\label{sec:deskValidation}
In the following, we provide a brief summary of the advantages and drawbacks of the application of artifact-based RE and the first version of AM4RRE with a focus on its applicability for managing legal knowledge and engineering-legal coordination.

\textbf{Summary of the main advantages} of AM4RRE that participants identified is as follows: inclusion of legal experts as stakeholders (RER1); representation of both engineering and legal concerns (LR1, LR3); facilitation of engineering-legal communication (RER1, LR1, LR2, LR3); explication of legal knowledge to a certain degree (RER1, LR1, LR2, LR3).

\begin{colorb}
"This [walkthrough] is the way I was taught to do legal interpretation at the University. Amazing that you can model it." (One of the focus group participants after a validation walk-through)
\end{colorb}

\begin{colorb}
"I think it's quite straightforward to map, to understand their connections and also to capture the legal parts of information. I think definitely it can help to get information that is necessary together and also understand the relations, how these information influence or impact aspects in our development life cycle." (RER1)
\end{colorb}

\textbf{Summary of the main drawbacks} of AM4RRE according to the participants are as follows: incapability to fully replace communication and interpretation (RER1, RER2, LR2, LR3); model can be complex to use (RER1, RER2, LR1); more flexibility can be required for different cases (LR1, LR2, LR3), unclear how variability will be addressed (RER2, LR2, LR3).

\begin{colorb}
"It can make it easier for the non-legal experts to stay compliant, but it will not compensate the interpretation of the law in general. The person who uses the model [needs] to explain what has to be done has to write it in the model in the right way." (LR1)
\end{colorb}

All legal researchers agreed that the model can \textit{explicate legal knowledge and partially compensate when an engineering team lacks such knowledge}. LR2 mentioned that AM4RRE embeds legal knowledge by reflecting on the structure of the law, some of its concepts, and how the law operates on a very abstract level. Yet, participants also mentioned that it can support legal interpretation only to a limited degree. Some reasons for that are the absence of a uniform use of some concepts even by legal scholars and practitioners (LR2), the complexity of the legal concepts, and the multiplicity of intricacies (LR1, LR2) (e.g., differentiation between regulatory act being in force and effect (LR2), implicit changes in legal interpretation due to economic or political conditions (LR1) or changes in sources used for interpretation (LR2)).

Legal researchers LR2 and LR3 also expressed concerns that ``something can get lost'' (LR2) in case of application of the model, the model can ``surrogate the actual law'', and the model won't be able to account for the uncertainty of interpretation of the law by courts (LR3). According to them, all these can result in non-compliance without the opportunity to identify it effectively. RER1, finally, also suggested that the model cannot fully replace communication between the roles.

Regarding the \textit{engineering-legal interaction}, RER1 corroborated the challenge as software engineers and legal experts usually work in isolation and present their results in a non-structured textual form. LR1 stated that the model can provide a communication framework for legal experts, who often work unstructured. LR2 further suggested that regulators and legal experts can use the model to understand how regulatory acts were interpreted in the regulatory RE process and identify the responsible roles.

\begin{colorb}
\textbf{Answer to RQ3:} Both RE and legal researchers find that the artifact-based approach captures legal knowledge on a high level and can serve as a basis for engineering-legal coordination. Aspects of the practical applicability of the model, including the degree to which it can \emph{efficiently} capture legal knowledge, need to be further validated in practical settings.
\end{colorb}

\section{Exemplary Application of AM4RRE}
This section briefly introduces aspects of our model's practical application. We introduce an exemplary use case and instantiation before reflecting on benefits and drawbacks (informed through our validation) to steer future work.

\subsection{Use Cases}
We envision two main use cases for the application of AM4RRE. Firstly, it is \textit{explication and management of legal knowledge} through the specification of the required legal concepts and their connections to requirements concepts. Detailed descriptions of legal concepts and their properties can help to \textit{compensate for the potential lack of legal experts}. Secondly, when integrating legal experts in a project, the artifact model can \textit{guide the roles in the independent specification of artifacts they are responsible for}. The \textit{engineering-legal coordination} is guided by the relationships in the artifact model. Next, an exemplary instantiation is shown to illustrate the use cases on a conceptual level (without its operationalization aspects).

\subsection{Exemplary instantiation of the AM4RRE content model}

\begin{figure}
\centerline{\includegraphics[width=1\linewidth]{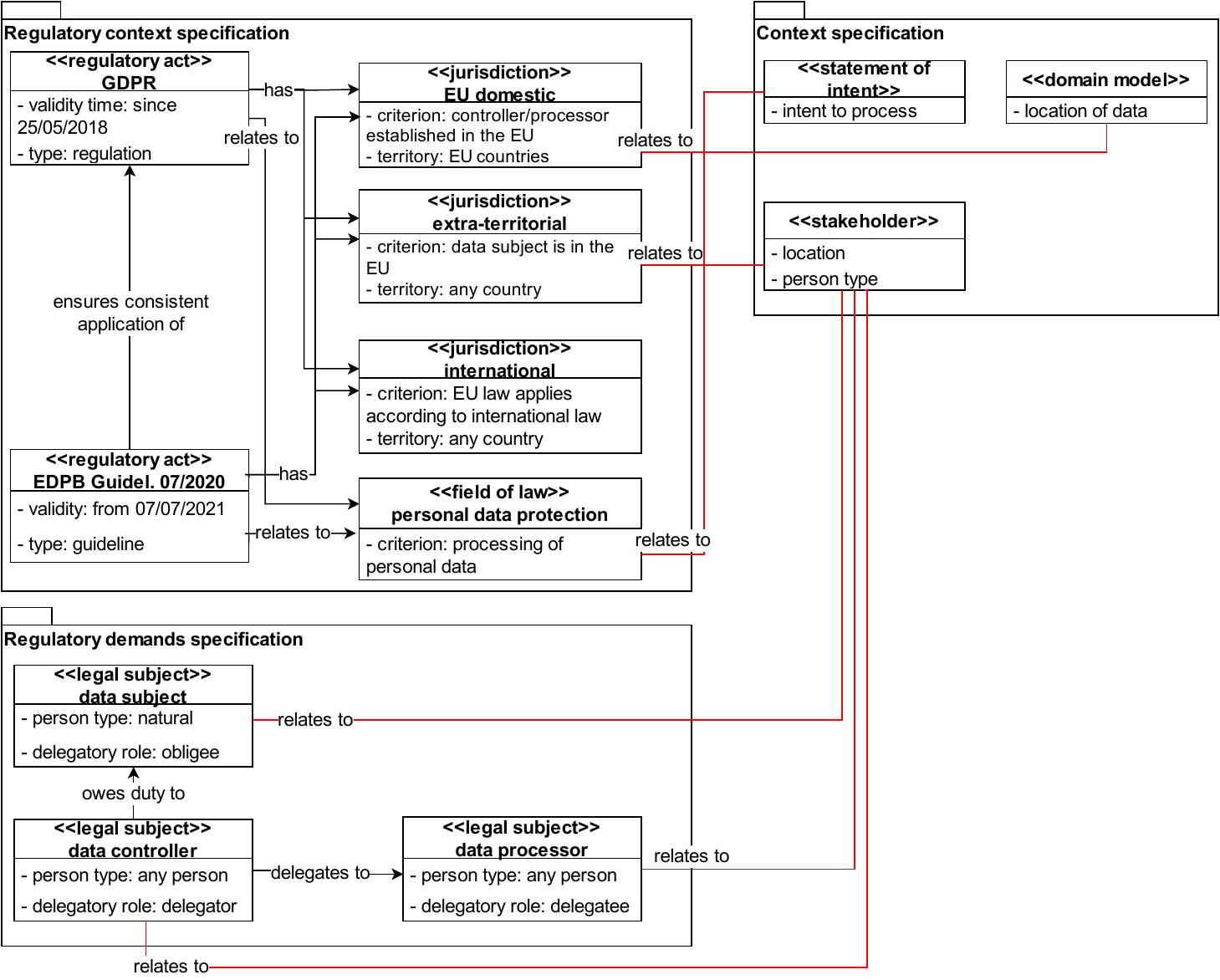}}
\caption{Exemplary instance of selected classes of the artifact content model for two regulatory acts: GDPR and EDPB Guidelines 07/2020.}
\label{fig:am4rreInstance}
\end{figure}

We provide the following example of the specification of selected classes and properties of the content model for compliance with personal data protection regulatory acts (see Figure~\ref{fig:am4rreInstance}). This example demonstrates how including legal concepts contributes to conducting legal interpretation as a part of regulatory RE and, hence, creates a basis for engineering-legal coordination. Due to space limitations, we use only selected legal concepts and their properties. We do not consider any particular instance of classes and concepts belonging to a software project (context, requirements, and system specification). We do this intentionally to demonstrate how regulatory context and demand specifications can be instantiated and later reused independently of particular software projects.

\textbf{Instantiation of legal concepts for Regulatory context specification}. In this example, we consider the instantiation of the following classes in the content model (C3.2): <<regulatory act>> (instances ``GDPR'', ``EDPB 07/2020''), <<jurisdiction>> (instances ``EU domestic'', ``extraterritorial'', ``international''), <<field of law>> (instance ``personal data protection''). The properties of the instances and the relationships between them determine the priority of processing regulatory acts ('regulation' has a higher force than 'guideline'). In our model, this implicit legal knowledge is made explicit by specifying the types of regulatory acts as properties of ``GDPR'' and ``EDPB 07/2020'' and by adding the relationship ``ensures consistent application of'' between GDPR and EDPB guidelines.
The two regulatory acts that we consider will be \textit{applicable} to a project if (1) ``location'' specified for an instance of <<stakeholder>> or ``location of data processor'' specified for an instance of <<domain model>> match ``criteria'' in <<jurisdiction>> (Regulatory context specification) and (2) ``intent to process personal data'' specified in <<statement of intent>> (Context specification) match the 'criteria' in <<field of law>> Regulatory context specification). The instantiation of all the classes of the content model by both the Legal expert (R2) and Domain expert (R3) and the establishment of relationships between them marks the achievement of Milestone 2 (Acceptance of regulatory context specification) and identification of applicable regulatory acts.

\textbf{Instantiation of legal concepts for Regulatory demands specification}. Next, regulatory demands contained in the applicable regulatory acts are specified by the Legal expert (R2). Here, we limit our example to the specification of one class of <<legal subject>> (instances ``data subject'', ``data controller'', and ``data processor''). Checking compliance of software systems to regulatory acts requires the differentiation between the ``data controller'' and ``data processor'' as it determines the applicable regulatory demands~\cite{amaral2023nlp}. For now, fully automatic differentiation between ``data controllers'' and ``data processors'' can be challenging and rely on human input~\cite{amaral2023nlp}. In practice, it is also important to understand that both the ``data controller'' and ``data processor'' act in the interests of the ``data subject'' (i.e., the ``data subject'' can execute their rights both in relation to the ``data controller'' and ``data processor''). This complex relationship can be explicated and captured with a legal concept of delegation (i.e., giving another person the responsibility of carrying out or executing certain actions)~\cite{saylor2012delegation}. Our model ensures the explication of delegation as a legal concept through (1) the specification of ``delegatory role'' as a property of instances of <<legal subject>>, (2) the specification of relationships of "owes duty to" between instances of <<legal subject>> having a value of ``delegatory role'' property 'delegator' and 'obligee' and relationship of "delegates to" between 'delegator' and 'delegatee'.
Specification of 'person type' property is important for further legal interpretation, which we model through the instantiation of relationships between regulatory demands specification and context specification (red lines in Figure~\ref{fig:am4rreInstance}). To establish the instances of <<legal subject>> class (in Regulatory demands specification C3.1.2) related to instances of <<stakeholder>> class (in Context specification C3.1.3) Legal expert will require specification of ``person type'' for <<stakeholder>>. As the ``data subject'' instance can only be a 'natural person' all instances of <<stakeholder>> not having a value of 'natural person' are not considered to be 'data subjects' from the legal perspective. This way, legal interpretation includes (1) iterative instantiating of legal concepts contained in a regulatory context and regulatory demands specifications and (2) establishing relationships to the instances of concepts in context, requirements, and system specification. This results in regulatory demands specification being completed and accepted (Milestone 3 is achieved). Specification of further regulatory requirements (Milestone 4) can be completed after the consistency of all the classes in context, requirements and system specification is established. We skip the achievement of this milestone in this paper.

\section{Future work}
The focus group results and conceptual validation provide a conceptual basis for our future work toward applying artifact-based RE for (1) effectively capturing and reflecting legal domain knowledge and (2) engineering-legal coordination.

Our current research results are of a conceptual nature as both focus group research and validation were conducted with researchers and, in a way, isolating requirements engineering and legal perspectives. We plan to extend our work and conduct a \textit{case study} to empirically identify the aspects of legal domain knowledge management and engineering-legal coordination relevant in practice. Such a case study will allow us to identify the practical usefulness of the artifact-based regulatory RE approach in practice. For now, we envision the following benefits of AM4RRE, which need further confirmation in future empirical research:
\begin{itemize}
    \item \textit{explication of legal knowledge}. Legal knowledge explicated in AM4RRE specification can be reused across further projects without or with only limited involvement of legal experts.
    \item \textit{explication of legal interpretation}. AM4RRE captures both legal concepts and RE concepts in their interconnection, which allows the legal interpretation process to be explicit and documented.
    \item \textit{facilitating completeness and consistency}. AM4RRE can potentially improve the completeness of requirements deriving from regulations (in relation to regulatory text) by providing a minimally required syntax for capturing legal concepts applied for the specification of applicable regulatory acts and their content. AM4RRE also aims to guide such consistency across different specifications (e.g., context, requirements, system specification) by prescribing relationships to adhere to.
    \item \textit{facilitating coordination}. Both the role model and the activity model (via milestones) can make explicit the responsibilities and concepts where coordination between engineers and legal experts is required.
\end{itemize}

Our future work on the operationalization of the model and its operational validation will address \textit{the limitations which were already identified during the conceptual validation}. These are as follows:
\begin{itemize}
	\item \textit{only basic guidance of legal interpretation and coordination}. Our current work presents a conceptual basis which also encompasses the opportunity to add further legal concepts required for the legal interpretation. Our future work will include extending the model to support its application in complex real-world scenarios.
	\item \textit{cumbersomeness and potential ineffectiveness of the model}. This limitation will be addressed by carefully designing the model operationalization methods. We consider that the application of checklists (what content to specify) can provide basic legal knowledge management and coordination, while templates and tool support (how to specify content) can help to address more complex regulatory RE situations.
	\item \textit{support for more flexibility and variability}. We plan to address this limitation with the development of \textit{tailoring profiles} to accommodate for differences in regulatory acts and software projects.
\end{itemize}

\section{Conclusion}\label{sec:conclusions}
In many cases, regulatory acts do not immediately specify software requirements. The derivation of software requirements from regulatory acts, which are valid from a legal perspective, should consider the legal interpretation process. This process is iterative and non-linear. It also involves the consideration of the regulatory act and its application in specific cases. It is driven by legal concepts well-known to legal experts but implicit or unknown to requirements engineers.

Capturing the elementary concepts and relationships in a legally relevant way and facilitating the interaction with legal experts remains challenging in regulatory RE. Nevertheless, ensuring the appropriate processing of regulatory acts in Software Engineering is essential. In this paper, we introduced an artifact-based regulatory RE approach and discussed how we developed it and how the validation is steering current development and evaluation tasks. As demonstrated, it has the potential to address many challenges by specifying legal concepts required for conducting legal interpretation as done in practice.
The results from our conceptual validation demonstrate how AM4RRE can support a structured interaction between engineers and legal experts.
Currently, we are enriching AM4RRE with further components that support operationalisation and application in industrial settings. One hope we associate with this manuscript is to demonstrate our conceptual (domain- and act-independent) model and expose it to critical discussions and feedback in the model-based RE community before continuing to develop tool-supported extensions and large-scale empirical evaluations in practice. Further, we want to motivate other scholars to join our endeavour while we enter the long-term instantiation phases for specific regulatory acts with model operationalization.

\section*{Data Availability}\label{sec:data}
We disclose auxiliary materials (validation questions, definitions of legal concepts used in AM4RRE, and Figures 1-4) \href{https://doi.org/10.5281/zenodo.11096716} {at Zenodo (DOI: 10.5281/zenodo.11096716).}

\section*{Acknowledgement}\label{sec:data}
This work was supported by the Bavarian Research Institute for Digital Transformation (bidt) through the "Coding Public Value" project and the KKS foundation through the S.E.R.T. Research Profile project at Blekinge Institute of Technology.

\bibliographystyle{IEEEtran}
\bibliography{IEEEabrv,bibliography}

\end{document}